\documentclass[10pt]{iopart}

\usepackage{iopams}
\usepackage{graphicx}
\usepackage{dcolumn}
\usepackage{bm}
\usepackage{wasysym}
\begin{document}


\title{Superconducting MoSi nanowires }

\author{J.~S.~Lehtinen, A. Kemppinen, E. Mykk\"{a}nen, M. Prunnila and A.~J.~Manninen } 
\address{$^1$ VTT Technical Research Centre of Finland Ltd, 02150 Espoo, Finland} 
\ead{janne.lehtinen@vtt.fi}

\begin{abstract} 
We have fabricated disordered superconducting nanowires of molybdenium silicide. A molybdenium nanowire is first deposited on top of silicon, and the alloy is formed by rapid thermal annealing. The method allows tuning of the crystal growth to optimise, e.g.,~the resistivity of the alloy for potential applications in quantum phase slip devices and superconducting nanowire single-photon detectors. The wires have effective diameters from 42 to 79 nm, enabling the observation of crossover from conventional superconductivity to regimes affected by thermal and quantum fluctuations. In the smallest diameter wire and at temperatures well below the superconducting critical temperature, we observe residual resistance and negative magnetoresistance, which can be considered as fingerprints of quantum phase slips.
\end{abstract}

\maketitle

\section{Introduction}

Highly disordered amorphous quasi-one-dimensional (1D) compound superconductors are interesting for the superconductor community, mainly for quantum phase slip (QPS) research~\cite{Harmans2005,Mooij2006,Hriscu2011a,Hriscu2011b,Lehtinen2012c,Hongisto2012,Astafiev2012,Kafanov2013,Peltonen2013,Arutyunov2016a} and superconducting nanowire single-photon detectors~\cite{Szypryt2016,Chandra2012,Banerjee2017}. Especially interesting are both materially and geometrically homogeneous 1D wires in which the finite resistivity in superconducting state can originate only from fluctuations of the phase of the order parameter. In comparison, in two-dimensional (2D) films the possible defects and fluctuations are shunted by adjacent superconducting channels. 

Development of modern fabrication techniques has enabled experimental investigations of size dependent fluctuations in 1D superconductors, and several QPS experiments with various materials MoGe~\cite{Bezryadin2000,Lau2001}, Al~\cite{Zgirski2005,Altomare2006}, Ti~\cite{Lehtinen2012a}, Nb~\cite{Cirillo2012,Zhao2016}, InO$_x$~\cite{Astafiev2012}, NbSi~\cite{Hongisto2012} and NbN~\cite{Peltonen2013,Arutyunov2016d} have  been made. The list is not comprehensive, but it gives an idea about the number of experimental studies so far. To obtain high fluctuation rate, the diameter of the nanowire must typically be below or at least comparable to the limits of modern e-beam lithography of about 10 to 20 nm. Fabrication of nanowire samples has utilized several unorthodox techniques, for example templating~\cite{Bezryadin2000,Lau2001}, electromigration~\cite{Baumans2016}, and ion milling~\cite{Zgirski2005,Lehtinen2012a}. The use of highly disordered and thus resistive materials has relaxed the size requirement but also made the fabrication more complex in sense of material homogeneity and available deposition techniques.  

Some of our initial objectives were to develop a fabrication routine that only utilizes techniques that are scalable on the wafer scale, and to achieve the aspect ratio between the height and the width of the cross section close to 1:1. A common problem of typical nanowire fabrication processes is that they use ultra thin films to achieve the required cross section and that increases relative inhomogeneity and surface area of the wire, making it more susceptible to the environment. This work progresses towards our future aim to develop a robust process which could be combined to the established silicon fabrication techniques to enable a complete silicon-based superconducting toolbox. 

As the material, we selected thermally formed molybdenium silicide (MoSi) which is well compatible with CMOS processes for quantum circuits~\cite{Zwanenburg2013}. In general, silicides have several ideal properties: tunable superconducting critical temperature $T_c$, possibility to tailor normal state resistivity from low to very high values, and proven high quality of the superconductor~\cite{Szypryt2016,Banerjee2017,Szypryt2015} when the device is fabricated on proper substrate and with diffusion barriers.

\section{Theory}

In quasi-one-dimensional nanowires the fluctuations of the phase of the superconducting order parameter cause a measurable resistance below $T_c$. The fluctuation governed superconductivity has been studied for more than fifty years. The concept of thermally activated phase slips (TAPS) was introduced by W. Little in 1967~\cite{Little1967} and they were observed experimentally a few years later~\cite{Lukens1970,Newbower1972}. According to Langer-Ambegaokar-Halperin-McCumber model~\cite{Langer1967,McCumber1968,McCumber1970}, the resistance below superconducting transition temperature is 
\begin{equation}
R_\mathrm{TAPS} = {\Omega \pi \hbar^2 \over 2e^2 k_B T} \exp(-{\delta F_0 \over k_B T}) \approx R_N  \exp(-{\delta F_0 \over k_B T}),
\end{equation}
where 
\begin{equation}
\delta F_0 \approx 2.2 {R_Q \xi(0)  \over R_N L} k_B T_c (1-T/T_c)^{3/2} 
\end{equation}
is the free energy barrier for phase slips valid at temperatures $T > 0.7~T_c$~\cite{Semenov2010}. Here, $\Omega$ is the attempt frequency, $R_Q = h / (2e)^2$ is the superconducting resistance quantum, $h$ and $\hbar$ are Planck and reduced Planck constants, respectively, $k_B$ is the Boltzmann constant, $e$ is the elementary charge, $L$ is the length of the wire and $\xi(T)$ is the temperature dependent superconducting coherence length. For experimental purposes the prefactor of $R_\mathrm{TAPS}$ can be adequately approximated with the low temperature normal state resistance $R_N$~\cite{Bezryadin2008}.

In the dirty limit $l \ll \xi_\mathrm{BCS}$, true for all our wires, the superconducting coherence length $\xi(T) = \sqrt{l \xi_\mathrm{BCS}(T)}$ can be estimated using the well-known Bardeen-Cooper-Schrieffer (BCS) relation $\xi_\mathrm{BCS}(T)= \hbar v_F / (\pi \Delta(T))$, where $\Delta(T)$ is the temperature dependent BCS energy gap, $v_F$ is the Fermi velocity and $l$ is the mean free path. This leaves product $l v_F$ as the only fitting parameter since the other parameters can be measured directly or calculated from the BCS theory. 

Thermally activated phase slips are of essence only when $T \approx T_c$, and they are suppressed exponentially when $T \ll T_c$.  Instead the QPS gives rise to finite resistance that can be observed down to $T = 0$. The model by Golubev and Zaikin~\cite{Zaikin1997,Golubev2001} has been successfully employed in several experiments~\cite{Zgirski2005,Altomare2006,Arutyunov2008,Bezryadin2008,Zgirski2008,Lehtinen2012a,Zhao2016,Arutyunov2016d} to explain observations of the residual resistance due to QPS well below $T_c$  in various materials. 

The model predicts the resistance
\begin{equation}
R_\mathrm{QPS} = b S_\mathrm{QPS}^2 \Delta(T) {L \over \xi(T)} \exp(-2 {S_\mathrm{QPS}}),
\end{equation}
where $S_\mathrm{QPS}=A (R_Q / R_N)(L/\xi(T))$ is the QPS action. The parameters $A \approx 1$ and $b \sim R_Q / \Delta(0)$ are constants, but cannot be precisely accounted within the model~\cite{Zaikin1997,Golubev2001,Arutyunov2008}. The model assumes that the wire is homogeneous. In real wires, there are always weak links, which localize the phase slips~\cite{Hriscu2011,Vanevic2012}. Yet for short wires this effect should be sufficiently small to be ignored when compared to inevitable uncertainty of the parameters $A$ and $b$. Also the model is applicable only for small QPS rates~\cite{Arutyunov2008,Lehtinen2012b}, but as far as we know, no microscopic theory exists for the case when QPS dominates the charge transport.  

The quantum phase slips can be considered as the magnetic counterpart for charge tunneling in a Josephson junction (JJ). Instead of tunnelling of Cooper pairs through an electrically insulating layer, magnetic vortices tunnel through the superconductor (magnetic insulator)~\cite{Mooij2006}. The duality to JJ physics has allowed mapping of the extensive JJ theory to QPS elements when proper energy scales and quantum conjugate variables of charge and phase are exchanged. The theory predicts insulating transition (Coulomb blockade) instead of superconducting transition for nanowires with sufficiently high QPS amplitude ~\cite{Hongisto2012,Lehtinen2012c,Mooij2015,Arutyunov2016c}.  This has enabled the realisation of several QPS based devices such as Cooper pair transistor~\cite{Hongisto2012,Arutyunov2016a}, flux transistor~\cite{Kafanov2013} and QPS qubit~\cite{Astafiev2012}. Even though the experiments have been promising, QPS elements are not extensively used in applications because of the lack of proper theoretical understanding of the phenomenon, complex fabrication methods, and so far low reproducibility of the elements.
 
\section{Fabrication and measurements}

The fabrication process of MoSi nanowires is depicted in Fig.~1. As a substrate we used float-zone~grown~$\langle$100$\rangle$ prime intrinsic 4 inch silicon wafers with very low impurity concentration and resistivity $\rho$ larger than  $ 10~$k$\Omega$m. The wafers were coated with about 200 nm thick positive e-beam resist AR-P 6200.6 (Fig.~1a). Electron beam writer with 100~kV acceleration voltage was used to pattern the structures with dose 250~$\mu$C/cm$^2$ and the resist was developed at room temperature with AR~600-546 for 2 minutes (Fig. 1b).  The native silicon oxide was removed by argon ion milling (Fig.~1c) prior to sputter deposition of 12~nm thick Mo-film (Fig.~1d). After the deposition the resist was chemically stripped in RM 600-71 within ultra-sonic bath (Fig.~1e). The room temperature resistivity of the deposited polycrystalline Mo-film was $26 \times 10^{-8}~\Omega$m. Taking into account the polycrystalline nature of the sputtered film, it agrees well with the bulk material value~\cite{Desai1984} of Mo, $6 \times10^{-8}~\Omega$m.

In annealing the Mo wire diffuses into silicon and forms the MoSi wire (Fig.~1f).  We used rapid thermal annealing in N$_2$ ambient with three parameter combinations: 400~$^\circ$C for 60~s, 600~$^\circ$C for 60~s and 900~$^\circ$C for 15~s. Samples fabricated with the latter two parameter sets suffered from high contact resistance between aluminum bonding wires and contact pads of the sample. When cooled to cryogenic temperatures the contact resistance increased to a degree that the samples could not be measured. Samples annealed at 400~$^\circ$C had no such problems and only those were used in the experiment.

\begin{figure}
	\includegraphics[width=\linewidth]{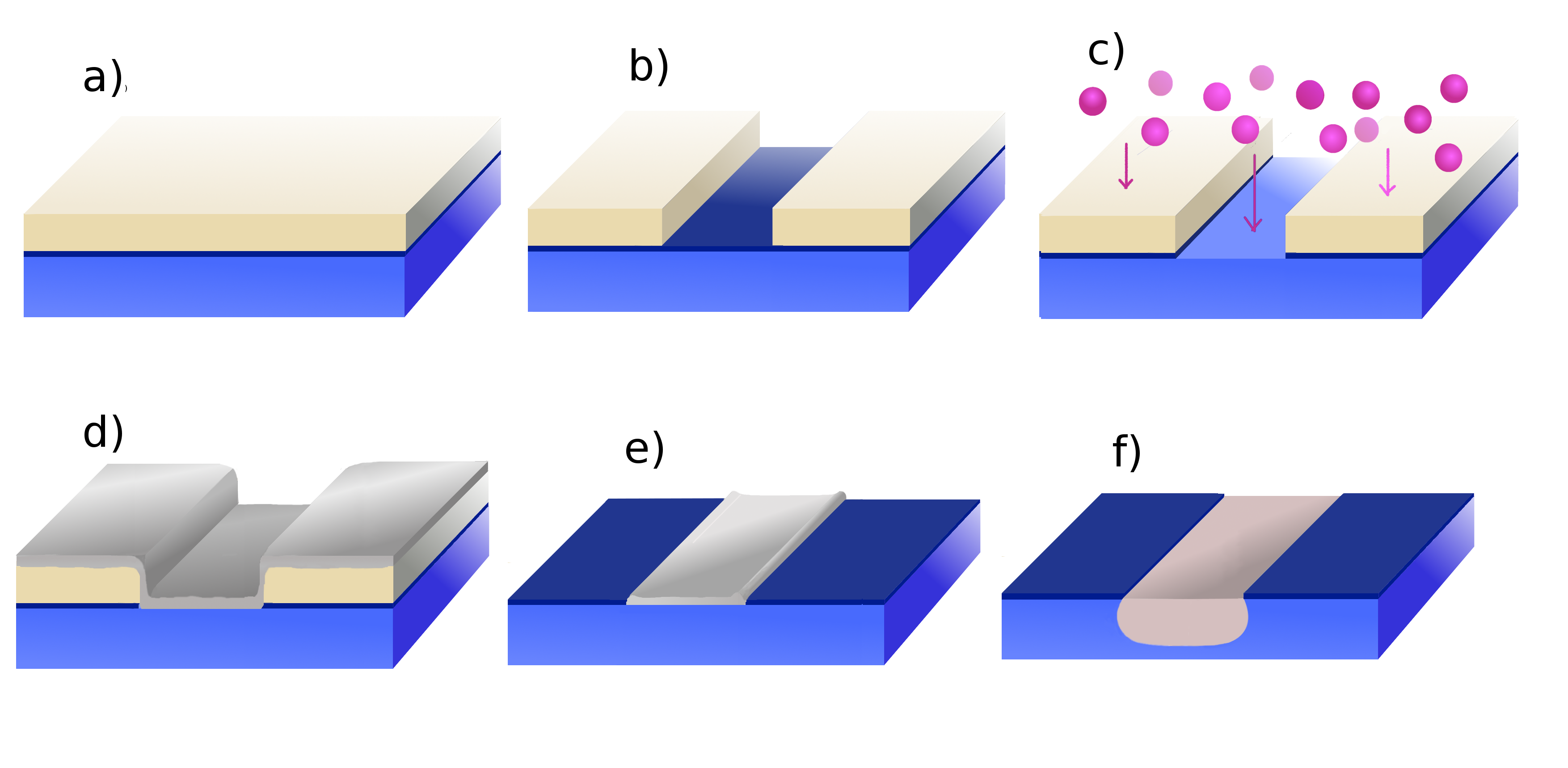}
  	\caption{Illustration of the process. a) Si-wafer with AR-P~6200.6 coating. b) Electron beam lithgraphy patterned resist. c) Ar etch prior to Mo deposition. d) Mo deposition on wafer.  e) Harsh lift-off in ultra-sonic bath to remove excess metal and side walls. f) Rapid thermal anneal diffuses the wire in the substrate leaving relatively smooth surface.}
  	\label{fig:process}
\end{figure}

The samples were measured in a dilution refrigerator at base temperature of about 15~mK. The measurement setup was the same that was used in Ref.~\cite{Lehtinen2017} with two nested microwave shields. Bronze powder filters~\cite{Lukashenko2008} and resistive miniature stainless steel coaxial cables (length 16 cm, diameter 0.33 mm) were used as high frequency filters~\cite{Zorin1995} in addition to typical lumped element RC-filtering. 


The lift-off method was used in this material research because of its simplicity. For future devices, the positive resist and lift-off technique can be replaced by negative resist (HSQ) and etching technique, respectively. Also diffusion barriers can be incorporated to the structure for a better control of the Mo diffusion in the Si substrate, for example by using silicon-on-insulator wafers. This has been shown to be  important for the suppression of the high-frequency losses in thermally formed PtSi resonators~\cite{Szypryt2016}. Yet it is not evident that the barriers are of advantage for the QPS applications. It is also likely that the contact resistance problems observed in samples annealed at $T \geqslant$~600~$^\circ$C can be solved by using two steps of lithography, where the contact pads are made of another material than the nanowire.

 \section{Results and discussion}    
 
\begin{figure}
	\includegraphics[width=\linewidth]{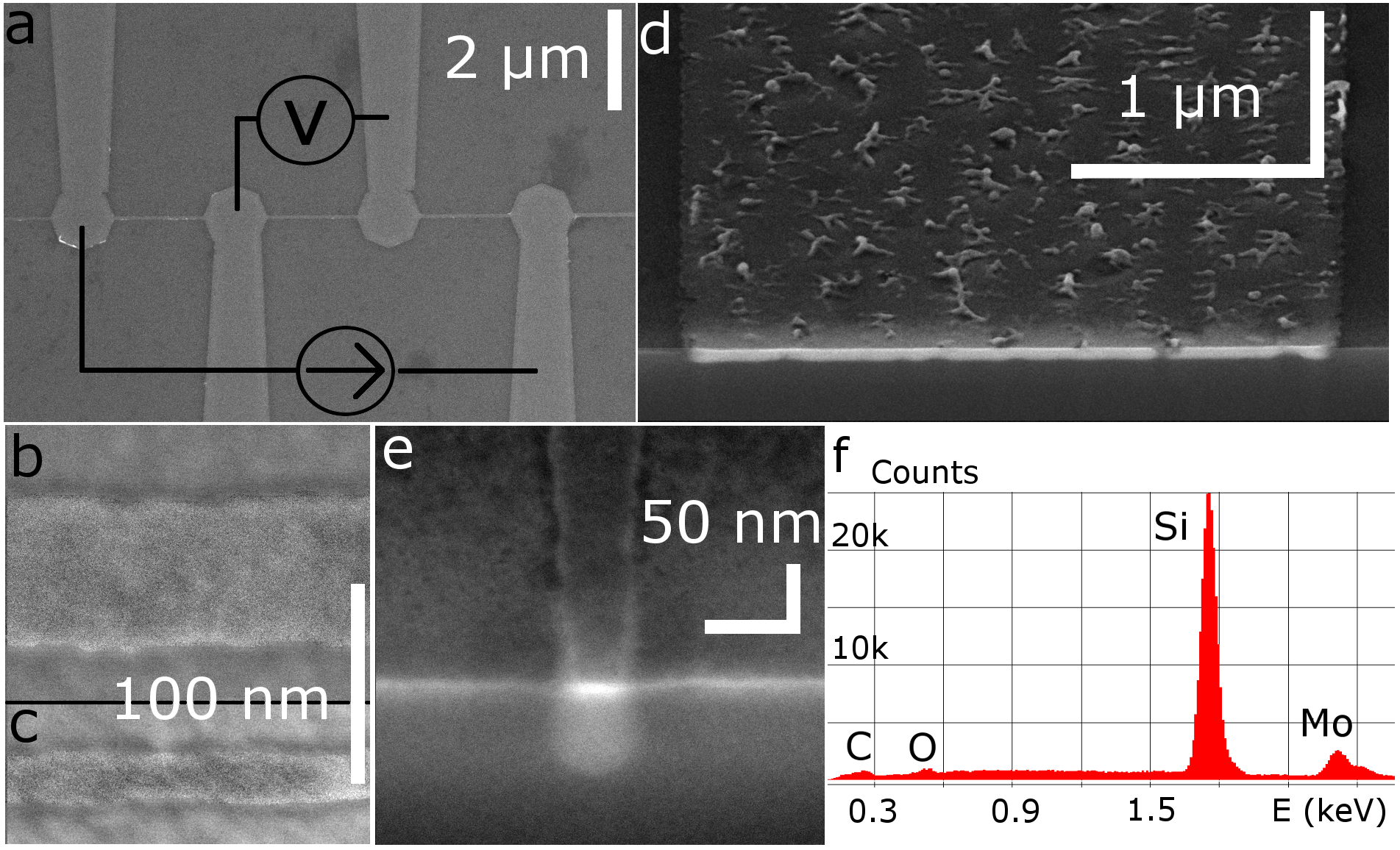}
  	\caption{a) Overview SEM image of the wire samples with the four probe measurement configuration. b) and c) SEM images of nanowires with widths 90 and 26 nm, respectively. d) and e) Cross-sectional SEM image of a micron scale electrode and a nanowire, respectively. f) EDX spectrum taken from the large wire d).}
  	\label{fig:SEMimages}
\end{figure}

A collection of scanning electron microscope (SEM) images of the MoSi nanowires are presented in Figs. \ref{fig:SEMimages}a-e. An overview of the sample and the four probe measurement configuration that was used in the measurements is shown in Fig.~\ref{fig:SEMimages}a. Two nanowires with widths 90 nm and 26 nm are presented in figures b) and c), respectively, showing high homogeneity of the lateral dimension. All the imaged nanowires were highly uniform and no obvious granular structure was observed in any of them. 

Cross sectional images of a micron scale electrode (Fig. \ref{fig:SEMimages}d) and a nanowire (Fig. \ref{fig:SEMimages}e) were prepared with focused Ga-ion beam etching and SEM imaging. Polymer residues from the resist or plasma etching are visible on large scale conductor, but all sub-micron structures are free from such contamination. The formation of MoSi is observed to be very uniform and smooth under the Si surface. Several nanowires were prepared and examined in a similar way. No irregularities or defects were observed in any of the measured wires. 

The cross sectional images show a clear boundary between Si and MoSi. The thickness of the formed MoSi film was $t = (70~\pm$~5)~nm. With this method, fabrication of almost arbitrary aspect ratios of the wire seems feasible. Our wires had the height to width ratio from about 1:2 to 3:1, the latter being hard to fabricate with any other means for these few-tens-of-nm wide wires.

We used energy-dispersive X-ray spectroscopy (EDX) for elemental analysis. An EDX spectrum measured from the sample of Fig.~\ref{fig:SEMimages}d is presented in Fig.~\ref{fig:SEMimages}f. The silicon content is overestimated because of the electrons penetrating deep into the substrate. The penetration depth of the incident e-beam that is about 1~$\mu$m determines the smallest volume possible for the analysis. Traces of oxygen and carbon from the native oxide layers and the polymer residues, respectively, are observed in addition to obvious Mo and Si, but no other contaminants were identified.

\begin{figure}
	\includegraphics[width=\linewidth]{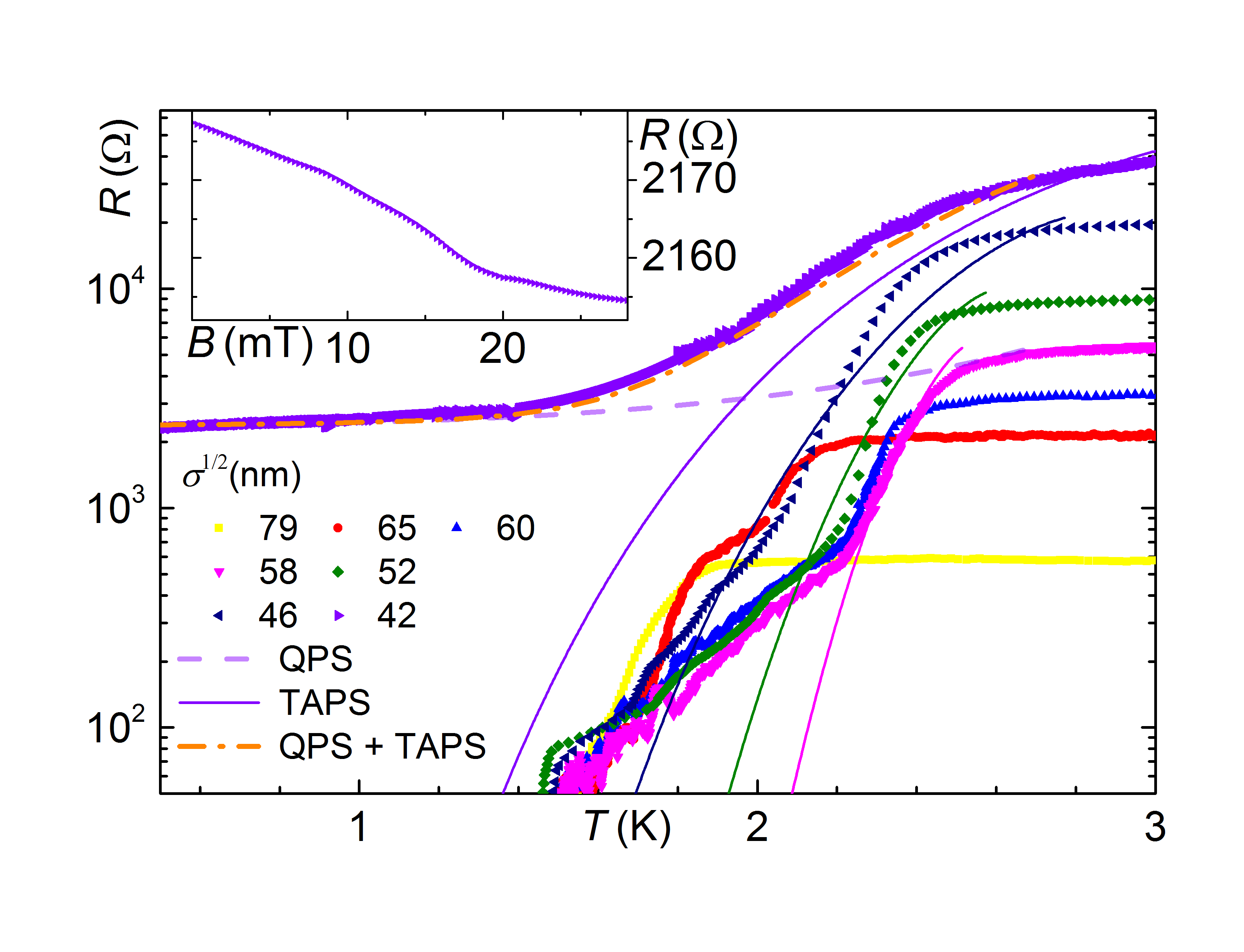}
  	\caption{Resistance versus temperature for seven nanowires with effective diameters indicated in the graph. Thermal (solid lines) and quantum fluctuations (dashed line) are fitted to the data of corresponding colour and their sum is marked with orange dash-dotted line. The inset shows resistance as a function of external magnetic field at $T = 15$~mK for the smallest wire. }
  	\label{fig:Tc}
\end{figure}

The resistance as a function of temperature for seven nanowires is presented in Fig.~\ref{fig:Tc}. Relatively sharp superconducting transition is observed in the widest nanowire with effective diameter $\sigma^{1/2} = (79~\pm~6)$~nm [width (90~$\pm$~3)~nm, thickness (70~$\pm$~5)~nm]. The wires with diameters from 46~to~65~nm have a resistive tail which has no size dependence. This is likely due to the contact electrodes fabricated from the wide MoSi film whose $T_c$ is lower than that of the nanowires. It can affect in two ways. First, the measured resistance has finite contribution from the electrode resistance because of the measurement configuration (Fig.~\ref{fig:SEMimages}a), and second, while being in normal state the contacts suppress the superconducting energy gap in the nanowire by inverse proximity effect. 

Theoretical TAPS fits (Eq.~(1)) for four of the smallest nanowires are shown in Fig.~3 with solid lines and with colour that corresponds the fitted data. The fits agree reasonably well with the measured transitions of the larger wires when the contribution of the resistive tail due to the contacts is neglected. The product of the mean free path and the Fermi velocity $l v_F = 2.2 \times 10^{-3}$~m$^2$/s was used as the only true fitting parameter. The best fit $l v_F$-product translates roughly to $v_F$ between 2$\times 10^5$~and~4$\times 10^5$ m/s and mean free path $l$ between 10~and~$5$~nm, which are reasonable. This leads to coherence length $\xi(T)$ from about 40~to~100~nm, which is comparable to the wire dimensions. The other parameters used in the fits were normal state resistances and superconducting critical temperatures; $R_{N1} = 5.65~$k$\Omega$ and $T_{c1} =2.54$~K (magenta), $R_{N2} = 9.89~$k$\Omega$ and $T_{c2} =2.60$~K (green), $R_{N3} = 22.1~$k$\Omega$ and $T_{c3} =2.80$~K (dark blue), and $R_{N4} = 55.8~$k$\Omega$ and $T_{c4} =3.40$~K (violet).

For the narrowest wire with diameter 42~nm the TAPS fit no longer even qualitatively agrees with the data. The contribution of the QPS (Eq.~(3)) is already relevant (dashed line). In the analysis we fixed the unimportant prefactor parameter of Eq.~(3) to value $b = R_Q / \Delta(0)$ and used $A$ as a fitting parameter. The best fit is obtained with $A = 0.54$ which is comparable to the values observed in other materials. Good agreement with the experiment is reached when both thermal and quantum fluctuations are taken into account (dash-dotted line) marking a crossover from thermal to quantum fluctuation influenced regime. 

Negative magneto resistance is observed in the thinnest wire (inset Fig.~\ref{fig:Tc}). The phenomenon is typical for wires in the regime of high QPS rate~\cite{Zgirski2008,Lehtinen2012b,Mitra2016,Baumans2016} and has been considered as a fingerprint of phase-slip-dominated dissipation~\cite{Vodolazov2012}. The effect has several possible explanations~\cite{Pesin2006,Vodolazov2007,Arutyunov2008,Vodolazov2012}, but no commonly accepted model exists yet. The most likely scenario, explaining all observations so far, is the enhancement of the retrapping current of the superconducting nanowire~\cite{Vodolazov2012,Baumans2016}.

\begin{figure}
	\includegraphics[width=\linewidth]{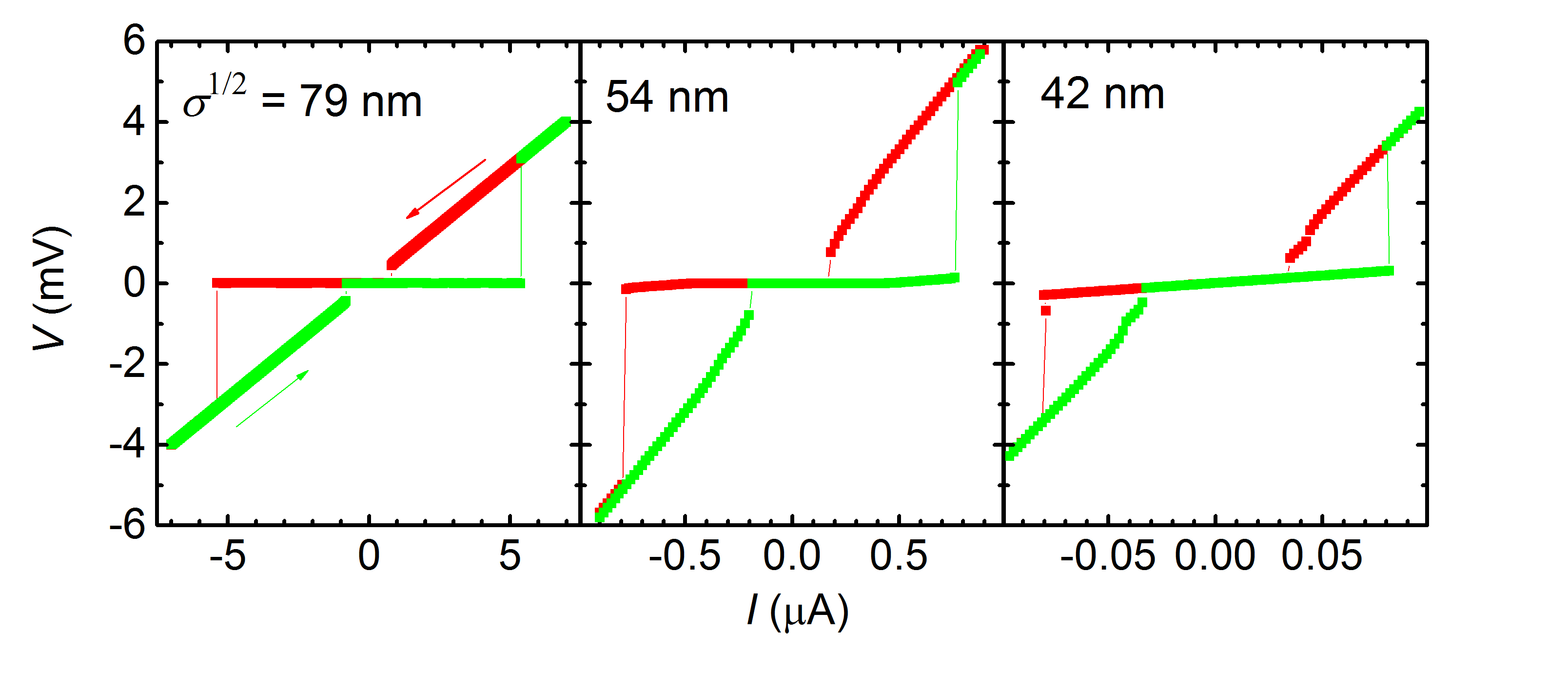}
  	\caption{IV-characteristics of three nanowires with effective diameters indicated in the graph measured at $T \approx 30$~mK. }
  	\label{fig:IVs}
\end{figure}

Current-voltage characteristics measured at $T \approx$~30~mK for three wires with effective diameters from 42 to 79~nm are presented in Fig.~\ref{fig:IVs}. The largest wire has typical characteristics of a superconductor with critical current density approximately 92~kA/cm$^2$ which is comparable to those observed in larger MoSi wires~\cite{Banerjee2017}. The second wire with diameter 54~nm has a significantly lower critical current density about 28~kA/cm$^2$. Finite resistance is observed already in the superconducting state before the switching,
probably originating from the fluctuations which become more pronounced as the increasing current lowers the free energy barrier that prevents the phase slips. The related dissipation drives the wire to the normal state much earlier than it would be in absence of fluctuations. The smallest wire with diameter 42~nm has a significantly suppressed critical current of about 4.5~kA/cm$^2$. Finite resistance, probably due to QPS, is observed already at sub-100~pA bias currents. 

\begin{figure}
	\includegraphics[width=\linewidth]{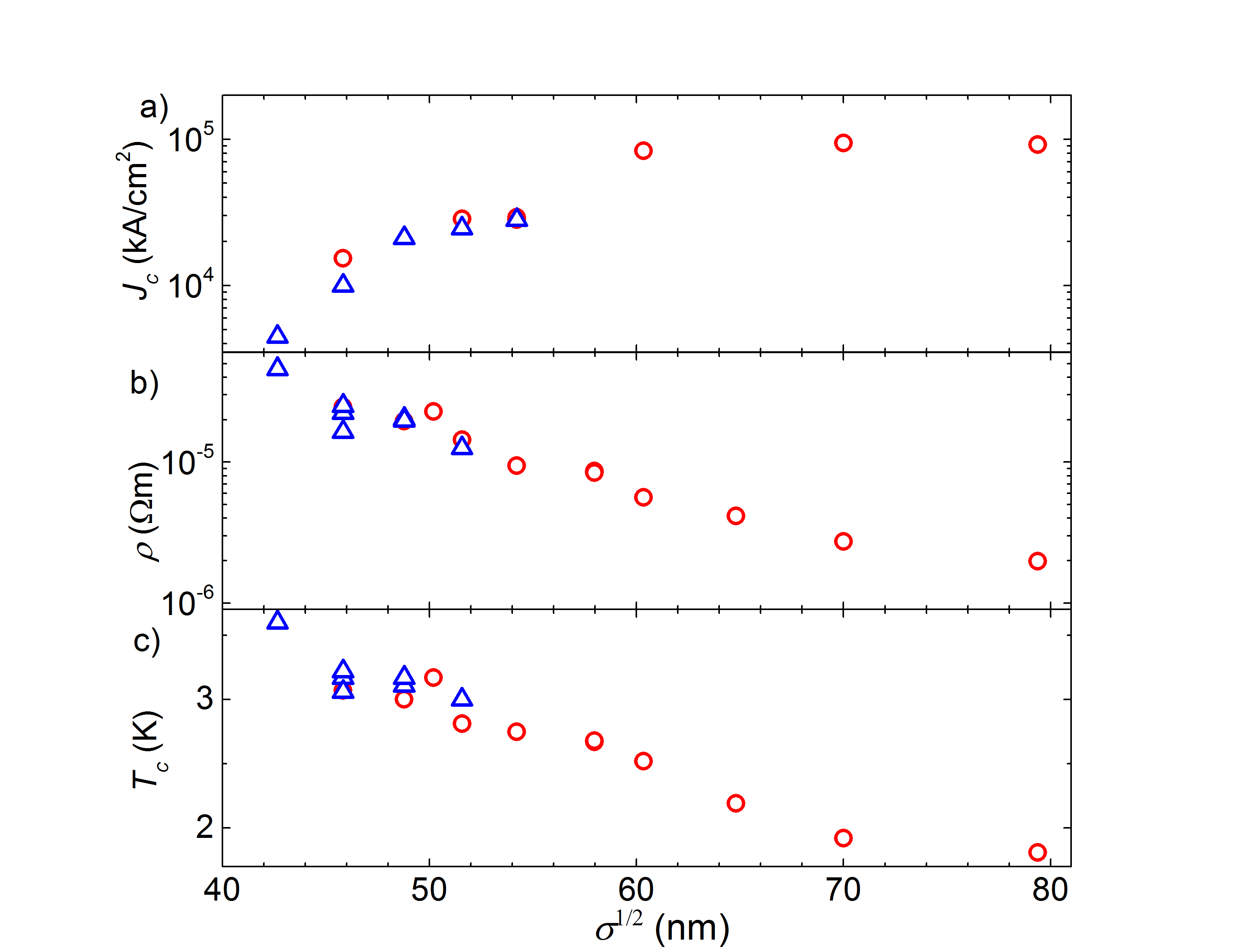}
  	\caption{a) Critical current density at 30 mK, b) low temperature normal state resistivity, and c) superconducting critical temperature of the nanowires as a function of the effective diameter. Samples annealed only at 400 $^\circ$C for 60 s are marked with red circles, and those annealed for the second time at 600~$^\circ$C for 30~s are marked with blue triangles.}
  	\label{fig:data}
\end{figure}

Critical current density data gathered from multiple wires are shown in Fig.~\ref{fig:data}a. For wires with $\sigma^{1/2} > 60$~nm the critical current density is constant, but for narrower wires it decreases with an exponential slope as a function of $\sigma^{1/2}$. This decrease of the critical current could be accounted by an exponentially increasing fluctuation rate as a function of diameter. 

The resistivity of the nanowires increases exponentially as the effective diameter is reduced (Fig.~\ref{fig:data}b). Yet the wires appear homogeneous on the cross sectional SEM images and the height of the formed wire seems to be almost independent on the width of the original Mo wire diameter. Additionally the width of the nanowire affects the critical temperature (Fig.~\ref{fig:data}c), which is here defined as the highest temperature where $R(T) < 0.9~R_N$. We have chosen a value that is close to the normal state resistance but still well resolved experimentally, so that it would be adequately applicable for the fluctuation affected wires, i.e., neglect the resistive transition tails. Previous experiments indicate that disorder increases the critical temperature from the value $T_c=1$~K of bulk crystalline MoSi up to $T_c=7.6$~K for amorphous MoSi films~\cite{Bosworth2015,Banerjee2017}. Thus it would be plausible that the grain growth at the surface of the nanowires is inhibited and more ordered crystalline structures are formed only within the MoSi-matrix. This leads to more amorphous films in the narrower wires where the surface area compared to the volume is larger than in the wider wires. This would explain both effects, increase of the normal state resistivity and increase of the critical temperature as the nanowire width becomes smaller. 

After the first measurement the samples were further annealed at 600~$^\circ$C for 60~s. The samples were then re-measured in a dilution refrigerator but no noticeable change either in resistivity or superconducting properties were observed as is shown in Fig. \ref{fig:data}, where the red circles are measurements after the first anneal and the blue triangles after the second anneal. We conclude that the sample tolerates thermal treatment after initial anneal because the Mo has formed a stable alloy with Si preventing further diffusion. 

The composition or structure of the nanowire is hard to measure directly since it would require high spatial resolution. In the SEM images we observed high geometric homogeneity of the thermally annealed wires and we expect the same to be true for material composition across the wire. Utilising text book values for the density and the atomic mass for both Mo and Si, and the dimensions of the Mo thin film as deposited and the annealed MoSi nanowire, we get a rough estimate that the composition MoSi$_x$ is silicon rich, i.e. $x \gtrsim 2$. The accuracy of the estimation is limited by the uncertainty of the deposited Mo film thickness as the deposition rate, which was used to determine the thickness, had been calibrated with significantly thicker films. 
 
The sheet resistances $R_{\square}$ of the measured nanowires were between 30~and~650~$\Omega$. It is expected that the sheet resistance should be closer to the superconducting resistance quantum $R_Q = 6.45~$k$\Omega$ to permit coherent phase slips~\cite{Astafiev2012}. This is well within achievable limits of our method for MoSi nanowires while maintaining sufficiently large dimensions for reproducible fabrication. First it is trivial to reduce the film thickness from 70 nm down to 20 nm. Based on the resistivity trend in
Fig.~\ref{fig:data}b, the reduction of effective cross section from 40 nm down to 20 nm should increase the resistivity by about one decade. When these two effects are combined the sheet resistance of 20~nm~$\times$~20~nm wire would be $R_{\square} \approx 20$~k$\Omega $, i.e. well above the required value. 

The QPS is not the only possible scenario for the observation of the resistive low-temperature tails, even though the QPS model fits the data with realistic parameters. Particularly in 2D systems, disorder and thickness driven dissipative metallic phases~\cite{Mason1999,Couedo2016} have been observed in various materials, but the origin of metallic states is still under debate. The occurrence of the metallic states and incorporated quasiparticle poisoning would also explain the negative magnetoresistance with improved quasiparticle diffusion to bulk electrodes. However, simple thermal arguments suggest that dissipative metallic phases are not the origin of the resistive low-temperature tail of our thinnest nanowire. Let us consider the case where the normal-metal part is long enough that no supercurrent can flow through it, as a short metallic section would be proximised by the adjacent superconductors, and would only act as a weak link that localises QPS~\cite{Vanevic2012}. In the IV curve of the smallest wire, see Fig. 4 rightmost panel, the highest measured current where the majority of the wire is still superconducting, is 80 nA. If the voltage of this measurement point, 0.3 mV, arose from a short normal-metal wire, it would dissipate the power of 24 pW. Yet only fraction of this power would heat the whole wire to normal state.

On the other hand, the QPS process does not have to be dissipative and produce heating~\cite{Mooij2006,Astafiev2012}. Dissipation in QPS circuits is still an open question and lacks proper microscopic model, but previous experiments have shown that phase slip centers can exhibit temperature-independent differential resistance as in our wires~\cite{Skocpol1974}. At low voltages this center carries a time-averaged supercurrent. In our smallest-diameter wire, observed low temperature resistance would suggest formation of a single spatially localized phase slip center with length of about 100 nm, which is between $\xi(T)$ and $2\xi(T)$.  



\section{Conclusions}
We have fabricated superconducting MoSi nanowires with standard electron beam lithography and rapid thermal annealing. The method enables wafer scale fabrication of highly resistive nanowires without obvious defects. 
We observe size dependent crossover from conventional superconductivity to transport influenced by thermal and quantum fluctuations. In the smallest diameter wire and at temperatures well below the superconducting critical temperature, we observe residual resistance and negative magnetoresistance, which based on the measured current-voltage characteristics cannot be explained by metallic weak links. Extending the process to yield nanowires with smaller diameter, and thus a higher QPS rate, requires only the deposition of a thinner Mo layer. These wires could be used to make QPS-based devices, for example photon, flux and charge detectors as well as thermometers and qubits. The first thermal anneal is crucial for the properties of the film, but after the initial anneal the wires did not suffer from heat treatment at 600~$^\circ$C. This is a critical property, when the wires are used in more complex structures, where several lithography cycles and thus use of elevated process temperatures is required. 

We would like to thank D. Golubev for discussions and acknowledge the Academy of Finland (Grants 288907 and 287768), Wihuri foundation and European Union's Horizon 2020 research and innovation programme (Grant Agreement 688539) for financial support. 


\begin{thebibliography}{10}

\bibitem{Harmans2005}
J.~E. Mooij and C.~J.~P.~M. Harmans.
\newblock Phase-slip flux qubits.
\newblock {\em New Journal of Physics}, 7(1):219, 2005.

\bibitem{Mooij2006}
J.E. Mooij and Yu.~V. Nazarov.
\newblock Superconducting nanowires as quantum phase-slip junctions.
\newblock {\em Nature Physics}, 2:169--172, 2006.

\bibitem{Hriscu2011a}
A.~M. Hriscu and Yu.~V. Nazarov.
\newblock Coulomb blockade due to quantum phase slips illustrated with devices.
\newblock {\em Phys. Rev. B}, 83:174511, 2011.

\bibitem{Hriscu2011b}
A.~M. Hriscu and Yu.~V. Nazarov.
\newblock Model of a proposed superconducting phase slip oscillator: A method
  for obtaining few-photon nonlinearities.
\newblock {\em Phys. Rev. Lett.}, 106:077004, 2011.

\bibitem{Lehtinen2012c}
J.~S. Lehtinen, K.~Zakharov, and K.~Yu. Arutyunov.
\newblock Coulomb blockade and {Bloch} oscillations in superconducting {Ti}
  nanowires.
\newblock {\em Phys. Rev. Lett.}, 109:187001, Oct 2012.

\bibitem{Hongisto2012}
T.~T. Hongisto and A.~B. Zorin.
\newblock Single-charge transistor based on the charge-phase duality of a
  superconducting nanowire circuit.
\newblock {\em Phys. Rev. Lett.}, 108:097001, 2012.

\bibitem{Astafiev2012}
O.~V. Astafiev, L.~B. Ioffe, S.~Kafanov, Yu.~A. Pashkin, K.~Yu. Arutyunov,
  D.~Shahar, O.~Cohen, and J.~S. Tsai.
\newblock Coherent quantum phase slip.
\newblock {\em Nature}, 484:355--358, 2012.

\bibitem{Kafanov2013}
S.~Kafanov and N.~M. Chtchelkatchev.
\newblock Single flux transistor: The controllable interplay of coherent
  quantum phase slip and flux quantization.
\newblock {\em Journal of Applied Physics}, 114(7), 2013.

\bibitem{Peltonen2013}
J.~T. Peltonen, O.~V. Astafiev, Yu.~P. Korneeva, B.~M. Voronov, A.~A. Korneev,
  I.~M. Charaev, A.~V. Semenov, G.~N. Golt'sman, L.~B. Ioffe, T.~M. Klapwijk,
  and J.~S. Tsai.
\newblock Coherent flux tunneling through nbn nanowires.
\newblock {\em Phys. Rev. B}, 88:220506, Dec 2013.

\bibitem{Arutyunov2016a}
K.~Yu. Arutyunov and J.~S. Lehtinen.
\newblock Junctionless cooper pair transistor.
\newblock {\em Physica C: Superconductivity and its Applications}, 533:158,
  2016.

\bibitem{Szypryt2016}
P.~Szypryt, B.~A. Mazin, G.~Ulbricht, B.~Bumble, S.~R. Meeker, C.~Bockstiegel,
  and A.~B. Walter.
\newblock High quality factor platinum silicide microwave kinetic inductance
  detectors.
\newblock {\em Applied Physics Letters}, 109(15):151102, 2016.

\bibitem{Chandra2012}
C.~M. Natarajan, M.~G. Tanner, and R.~H. Hadfield.
\newblock Superconducting nanowire single-photon detectors: physics and
  applications.
\newblock {\em Superconductor Science and Technology}, 25(6):063001, 2012.

\bibitem{Banerjee2017}
A. Banerjee, L.~J Baker, A. Doye, M. Nord, R.~M. Heath,
  K. Erotokritou, D. Bosworth, Z.~H Barber, I. MacLaren, and
  R.~H Hadfield.
\newblock Characterisation of amorphous molybdenum silicide (MoSi)
  superconducting thin films and nanowires.
\newblock {\em Superconductor Science and Technology}, 30(8):084010, 2017.

\bibitem{Bezryadin2000}
A.~Bezryadin, C.~N. Lau, and M.~Tinkham.
\newblock Quantum suppression of superductivity in ultrathin nanowires.
\newblock {\em Nature}, 404:971--974, 2000.

\bibitem{Lau2001}
C.~N. Lau, N.~Markovic, M.~Bockrath, A.~Bezryadin, and M.~Tinkham.
\newblock Quantum phase slips in superconducting nanowires.
\newblock {\em Phys. Rev. Lett.}, 87:217003, Nov 2001.

\bibitem{Zgirski2005}
M.~Zgirski, K.-P. Riikonen, V.~Touboltsev, and K.~Arutyunov.
\newblock Size dependent breakdown of superconductivity in ultranarrow
  nanowires.
\newblock {\em Nano Letters}, 5(6):1029--1033, 2005.

\bibitem{Altomare2006}
F.~Altomare, A.~M. Chang, M.~R. Melloch, Y.~Hong, and C.~W. Tu.
\newblock Evidence for macroscopic quantum tunneling of phase slips in long
  one-dimensional superconducting {Al} wires.
\newblock {\em Phys. Rev. Lett.}, 97:017001, 2006.

\bibitem{Lehtinen2012a}
J.~S. Lehtinen, T.~Sajavaara, K.~Yu. Arutyunov, M.~Yu. Presnjakov, and A.~L.
  Vasiliev.
\newblock Evidence of quantum phase slip effect in titanium nanowires.
\newblock {\em Phys. Rev. B}, 85:094508, 2012.

\bibitem{Cirillo2012}
C.~Cirillo, M.~Trezza, F.~Chiarella, A.~Vecchione, V.~P. Bondarenko, S.~L.
  Prischepa, and C.~Attanasio.
\newblock Quantum phase slips in superconducting nb nanowire networks deposited
  on self-assembled si templates.
\newblock {\em Applied Physics Letters}, 101(17):172601, 2012.

\bibitem{Zhao2016}
W. Zhao, X. Liu, and M.~H.~W. Chan.
\newblock Quantum phase slips in 6 mm long niobium nanowire.
\newblock {\em Nano Letters}, 16(2):1173--1178, 2016.
\newblock PMID: 26788964.

\bibitem{Arutyunov2016d}
K~Yu Arutyunov, A~Ramos-Álvarez, A~V Semenov, Yu~P Korneeva, P~P An, A~A
  Korneev, A~Murphy, A~Bezryadin, and G~N Gol’tsman.
\newblock Superconductivity in highly disordered NbN nanowires.
\newblock {\em Nanotechnology}, 27(47):47LT02, 2016.

\bibitem{Baumans2016}
X.~D.~A. Baumans, D. Cerbu, O.-A. Adami, V.~S.
  Zharinov, N. Verellen, G. Papari, J.~E. Scheerder, G. Zhang,
  V.~V. Moshchalkov, A.~V. Silhanek, and J. Van~de Vondel.
\newblock Thermal and quantum depletion of superconductivity in narrow
  junctions created by controlled electromigration.
  \newblock {\em Nat. Comm.}, 7:10560, Feb 2016.

\bibitem{Zwanenburg2013}
F.~A. Zwanenburg, A.~S. Dzurak, A. Morello, M.~Y. Simmons,
  L.~C.~L. Hollenberg, G. Klimeck, S. Rogge, S.~N. Coppersmith,
  and M.~A. Eriksson.
\newblock Silicon quantum electronics.
\newblock {\em Rev. Mod. Phys.}, 85:961--1019, Jul 2013.

\bibitem{Szypryt2015}
P.~Szypryt, B.~A. Mazin, B.~Bumble, H.~G. Leduc, and L.~Baker.
\newblock Ultraviolet, optical, and near-ir microwave kinetic inductance
  detector materials developments.
\newblock {\em IEEE Transactions on Applied Superconductivity}, 25(3):1--4,
  June 2015.

\bibitem{Little1967}
W.~A. Little.
\newblock Decay of persistent currents in small superconductors.
\newblock {\em Phys. Rev.}, 156:396--403, Apr 1967.

\bibitem{Lukens1970}
J.~E. Lukens, R.~J. Warburton, and W.~W. Webb.
\newblock Onset of quantized thermal fluctuations in "one-dimensional"
  superconductors.
\newblock {\em Phys. Rev. Lett.}, 25:1180--1184, Oct 1970.

\bibitem{Newbower1972}
R.~S. Newbower, M.~R. Beasley, and M.~Tinkham.
\newblock Fluctuation effects on the superconducting transition of tin whisker
  crystals.
\newblock {\em Phys. Rev. B}, 5:864--868, Feb 1972.

\bibitem{Langer1967}
J.~S. Langer and V. Ambegaokar.
\newblock Intrinsic resistive transition in narrow superconducting channels.
\newblock {\em Phys. Rev.}, 164:498--510, Dec 1967.

\bibitem{McCumber1968}
D.~E. McCumber.
\newblock Intrinsic resistive transition in thin superconducting wires driven
  from current sources.
\newblock {\em Phys. Rev.}, 172:427--429, Aug 1968.

\bibitem{McCumber1970}
D.~E. McCumber and B.~I. Halperin.
\newblock Time scale of intrinsic resistive fluctuations in thin
  superconducting wires.
\newblock {\em Phys. Rev. B}, 1:1054--1070, Feb 1970.

\bibitem{Semenov2010}
A.~V. Semenov, P.~A. Krutitskii, and I.~A. Devyatov.
\newblock Microscopic theory of phase slip in a narrow durty superconducting
  strip.
\newblock {\em JETP Letters}, 92(11):762--766, Dec 2010.

\bibitem{Bezryadin2008}
A. Bezryadin.
\newblock Quantum suppression of superconductivity in nanowires.
\newblock {\em Journal of Physics: Condensed Matter}, 20(4):043202, 2008.

\bibitem{Zaikin1997}
A.~D. Zaikin, D.~S. Golubev, A. van Otterlo, and G.~T.
  Zim\'anyi.
\newblock Quantum phase slips and transport in ultrathin superconducting wires.
\newblock {\em Phys. Rev. Lett.}, 78:1552--1555, Feb 1997.

\bibitem{Golubev2001}
D.~S. Golubev and A.~D. Zaikin.
\newblock Quantum tunneling of the order parameter in superconducting
  nanowires.
\newblock {\em Phys. Rev. B}, 64:014504, Jun 2001.

\bibitem{Arutyunov2008}
K.~Yu. Arutyunov, D.~S. Golubev, and A.~D. Zaikin.
\newblock Superconductivity in one dimension.
\newblock {\em Physics Reports}, 464(1):1 -- 70, 2008.

\bibitem{Zgirski2008}
M.~Zgirski, K.-P. Riikonen, V.~Touboltsev, and K.~Yu. Arutyunov.
\newblock Quantum fluctuations in ultranarrow superconducting aluminum
  nanowires.
\newblock {\em Phys. Rev. B}, 77:054508, Feb 2008.

\bibitem{Lehtinen2012b}
J~S Lehtinen and K~Yu Arutyunov.
\newblock The quantum phase slip phenomenon in superconducting nanowires with a
  low-ohmic environment.
\newblock {\em Superconductor Science and Technology}, 25(12):124007, 2012.

\bibitem{Hriscu2011}
  Hriscu, A. M. and Nazarov, Yu. V.
  \newblock Coulomb blockade due to quantum phase slips illustrated with devices
  \newblock {\em Phys. Rev. B}, 174511, 2011.

\bibitem{Vanevic2012}
  M. Vanevi\ifmmode \acute{c}\else \'{c}\fi{} and Yu.~V. Nazarov
  \newblock Quantum Phase Slips in Superconducting Wires with Weak Inhomogeneities
  \newblock {\em Phys. Rev. Lett.}, 187002, May 2012.


\bibitem{Mooij2015}
J~E Mooij, G~Sch\''{o}n, A~Shnirman, T~Fuse, C~J P~M Harmans, H~Rotzinger, and A~H
  Verbruggen.
\newblock Superconductor–insulator transition in nanowires and nanowire
  arrays.
\newblock {\em New Journal of Physics}, 17(3):033006, 2015.

\bibitem{Arutyunov2016c}
K.~Yu. Arutyunov, J.~S. Lehtinen, and T.~Rantala.
\newblock The quantum phase slip phenomenon in superconducting nanowires with
  high-impedance environment.
\newblock {\em Journal of Superconductivity and Novel Magnetism},
  29(3):569--572, 2016.

\bibitem{Desai1984}
P.~D. Desai, T.~K. Chu, H.~M. James, and C.~Y. Ho.
\newblock Electrical resistivity of selected elements.
\newblock {\em Journal of Physical and Chemical Reference Data},
  13(4):1069--1096, 1984.

\bibitem{Lehtinen2017}
J.~S. {Lehtinen}, E.~{Mykk{\"a}nen}, A.~{Kemppinen}, D.~{Golubev}, S.~V.
  {Lotkhov}, and A.~J. {Manninen}.
\newblock Characterizing superconducting filters using residual microwave
  background.
\newblock {\em Superconductor Science and Technology}, 30(5):055006, 2017.

\bibitem{Lukashenko2008}
A.~Lukashenko and A.~V. Ustinov.
\newblock Improved powder filters for qubit measurements.
\newblock {\em Rev. Sci. Instrum.}, 79:014701, 2008.

\bibitem{Zorin1995}
A.~B. Zorin.
\newblock The thermocoax cable as the microwave frequency filter for single
  electron circuits.
\newblock {\em Review of Scientific Instruments}, 66(8):4296--4300, 1995.

\bibitem{Mitra2016}
S. Mitra, G.~C. Tewari, D. Mahalu, and D. Shahar.
\newblock Negative magnetoresistance in amorphous indium oxide wires.
\newblock {Scientific Reports}, 6:37687, Nov 2016.

\bibitem{Vodolazov2012}
D.~Y. Vodolazov and F.~M. Peeters.
\newblock Enhancement of the retrapping current of superconducting microbridges
  of finite length.
\newblock {\em Phys. Rev. B}, 85:024508, Jan 2012.

\bibitem{Pesin2006}
D.~A. Pesin and A.~V. Andreev.
\newblock Suppression of superconductivity in disordered interacting wires.
\newblock {\em Phys. Rev. Lett.}, 97:117001, Sep 2006.

\bibitem{Vodolazov2007}
D.~Y. Vodolazov.
\newblock Negative magnetoresistance and phase slip process in superconducting
  nanowires.
\newblock {\em Phys. Rev. B}, 75:184517, May 2007.

\bibitem{Bosworth2015}
D.~Bosworth, S.-L. Sahonta, R.~H. Hadfield, and Z.~H. Barber.
\newblock Amorphous molybdenum silicon superconducting thin films.
\newblock {\em AIP Advances}, 5(8):087106, 2015.

\bibitem{Mason1999}
D.~Bosworth, S.-L. Sahonta, R.~H. Hadfield, and Z.~H. Barber.
\newblock Dissipation Effects on the Superconductor-Insulator Transition in 2D Superconductors
\newblock {\em Phys. Rev. Lett.}, 82:5341--5344, 1999.

\bibitem{Mason1999}
N.~Mason and A.~Kapitulnik
\newblock Dissipation Effects on the Superconductor-Insulator Transition in 2D Superconductors
\newblock {\em Phys. Rev. Lett.}, 82:5341--5344, 1999.

\bibitem{Couedo2016}
F.~Cou{\"e}do, O.~Crauste, A.~A.~Drillien, V.~Humbert, L.~Berg{\'e}, C.~A.~Marrache-Kikuchi, and L.~Dumoulin
\newblock Dissipative phases across the superconductor-to-insulator transition
\newblock {\em Scientific Reports}, 6:35834, 2016.

\bibitem{Skocpol1974}
W.~J.~Skocpol, M.~R.~Beasley and M.~Tinkham
\newblock {\em J. Low Temp. Phys.} 16:145, 1974.


\end{thebibliography}

\end{document}